\documentclass[12]{article}
\usepackage{graphics}

\newcommand{\beq}{\begin{equation}}
\newcommand{\eeq}{\end{equation}}
\newcommand{\Al}{\alpha (X)}
\newcommand{\al}{\alpha (x)}
\newcommand{\all}{\alpha}
\newcommand{\st}{\star }

\newcommand{\de}{\delta _{\al}}
\newcommand{\pd}{\partial}
\newcommand{\dd}[2]{\frac {\partial #1}{\partial #2}}
\newcommand{\lb}{\left(}
\newcommand{\rb}{\right)}
\newcommand{\beqs}{\begin{eqnarray}}
\newcommand{\eeqs}{\end{eqnarray}}
\newcommand{\half}{\frac{1}{2}}
\newcommand{\ab}{\theta (x)}
\newcommand{\dee}{\delta _{\epsilon}}
\newcommand{\chose}[2]{\left( \begin{array}{c} #1 \\ #2 \end{array} 
\right)}
\newcommand{\ovk}{\frac{i}{\kappa }}
\begin{document}
\bibliographystyle{h-physrev}
\input{epsf}

\title{Gauge Theories on Open Lie Algebra Non-commutative Spaces}
\author{A.Agarwal\thanks{abhishek@pas.rochester.edu} \\
\and
 L.Akant\thanks{akant@pas.rochester.edu} \\
University of Rochester. Dept of Physics and Astronomy. \\
Rochester. NY - 14627}
\maketitle

\begin{abstract}
It is shown that non-commutative spaces, which are quotients of 
associative algebras by ideals generated by highly non-linear 
relations of a particular type, admit extremely simple formulae 
for deformed or star products. Explicit construction of these star 
products is carried 
out. Quantum gauge theories are formulated on these spaces, and the 
Seiberg-Witten map is worked out in detail. 
\end{abstract}

{\bf Notation:} We shall use capital letters $X^i$ and $P^i$ to denote 
non-commuting 
coordinates. Functions of non-commuting coordinates will also be denoted 
by capital letters, e.g. $F(X,P)$. The corresponding commuting coordinates 
and their functions will be denoted by lower case letters, 
e.g. $x^i$, $p^i$ and $f(x,p)$.
\section{Introduction}
In the present paper, we consider a large class of  `open Lie algebras' as 
examples of non-commutative spaces. We use the term `open Lie 
algebras' to mean associative algebras where the defining relations
among the generators are consistent with the Jacobi identity but the
commutators of generators do not necessarily close on the set of
generators. 
In the open Lie algebras considered here (\ref{poisson1}), the 
commutators of the generators are allowed to be arbitrary functions 
of the generators, along with some restrictions which we shall elaborate 
on later in this section. 
We find that a 
judicious choice of ordering on these non-commutative spaces 
provides one with a surprisingly simple description of deformed  products. 
This allows for a detailed formulation of gauge theories on these 
non-commutative spaces. We do that following the method outlined by Madore 
et al\cite{Madoreetal}, and establish the Seiberg-Witten map 
\cite{Seiberg-Witten}. The 
organization of 
the paper 
is as follows. The rest of this section is devoted to the elaboration of 
our motivation. In the next section we work out the star product. The 
final section is devoted to the formulation of gauge theories and studying 
the Seiberg-Witten map.

The study of quantum field theories on non-commutative spaces (for 
reviews of the subject see \cite{ncreview1, ncreview2}) is made 
possible
largely due to success of the paradigm of deformation-quantization 
\cite{Kontsevich, Moyal, Bayenetal, Madoreetal, Seiberg-Witten}. The 
pioneering work in this field allows us to translate the problem of 
understanding functions on spaces with operator valued or non-commuting 
coordinates \cite{connesbook} into statements about functions of commuting 
variables with 
deformed (star, $\st $) products. For spaces, where the non-commutativity 
is in some sense simple, the isomorphism between functions of 
non-commuting variables  and functions of ordinary commuting variables 
assumes a particularly transparent form because of the availability of 
simple expressions for star products. The classic 
example of this is the study of a space which is the algebra formed by  
generators `$X^i$' obeying  
canonical 
commutation relations,
\beq
[X^i,X^j]=i\theta ^{ij}, \label{canonical}
\eeq
where $\theta ^{ij}$ is a constant, real, anti-symmetric matrix. For these 
spaces, the star product of functions $f$ and $g$ of ordinary 
variables $x^i$, can be expressed by the Groenwold-Moyal 
\cite{Moyal,groenwold} 
formula;
\beq
f(x) \st g(x) = f(x)\exp \lb \frac{i}{2}\stackrel{\leftarrow}{\pd 
_i}\theta 
^{ij}\stackrel{\rightarrow}{\pd _j}\rb g(x). \label{moyal}
\eeq
This deformed product corresponds to symmetric ordering or Weyl 
quantization. 

A measure of the degree of non-commutativity of a space, thought of as an 
abstract algebra, is encoded in the commutators of the generators. In 
some sense, the algebra with the relations (\ref{canonical}) forms a space 
that is `barely non-commutative'. That is so because (\ref{canonical}) 
defines a Lie-algebra, whose non-commutative part (the matrix $\theta ^ 
{ij}$) lies in the commutant of the algebra. At the next level of 
non-commutativity are Lie-algebras with non-vanishing structure constants. 
Analysis of quantum gauge theories on Lie-algebra non-commutative spaces, 
using Weyl quantization, was carried out by Madore et al\cite{Madoreetal}, 
and 
as one might expect, the formulae for star products in such spaces do not 
admit simple closed forms like (\ref{moyal}).
The ground breaking work of Kontsevich \cite{Kontsevich} and Cattaneo et 
al\cite{Catfel}, tells us how to 
extend Weyl quantization to generic non-commutative spaces, where the 
commutators of the coordinates are not constants, but arbitrary 
functions 
of the coordinates.
\beq
[X^i,X^j] = \theta ^{ij}(X) \label{generic}.
\eeq
Compared to (\ref{moyal}),
the formulae for the deformed products for generic non-commutative 
spaces (\ref{generic}) are much more involved, and for purposes of 
doing explicit 
computations of the sort required in the analysis of quantum field 
theories, it is certainly worthwhile to look for simplifications of the 
Kontsevich formula\cite{Kontsevich}. What is surprising is that for 
certain non-commutative 
spaces, whose generators satisfy non-linear relations, there do exist 
surprisingly explicit formulae for star products.

Deformed products of functions of commuting variables correspond to 
isomorphisms between functions of ordinary commuting variables and those 
of the corresponding non-commuting ones. The isomorphisms are obtained by 
specifying a choice of ordering for the functions of non-commuting 
variables. Hence some room for 
the simplification of the expressions for the deformed products is 
provided by our freedom in 
the choice of ordering. Moreover, all deformations of a particular 
commutative algebra of functions, obtained by different ordering 
prescriptions are equivalent up to 
cohomological issues \cite{Bayenetal, BakasKakas}.  
In the case of finite dimensional non-commutative spaces, 
which can be coordinatized by global coordinates, the subtle 
cohomological issues do not matter. In these cases, the 
cohomological 
equivalence of  orderings translates into a true equivalence, and  one is  
indeed free to choose a prescription of ordering at one's convenience. 
For 
the simplest case(\ref{canonical}), the star products are very explicit 
for almost any reasonable choice of ordering \cite{WolfAgarwal, Dunne}. 
For a geometric construction of (\ref{moyal}) and a review of some star 
products 
corresponding to various orderings in (\ref{canonical}) we refer to 
\cite{zachos1, zachos2}.

The non-commutative spaces that we consider in the present paper are 
associative algebras generated by a 
finite number of generators (modulo certain non-linear relations). Hence 
the 
argument for choosing a preferred ordering does apply here. We coordinatize 
the space by two sets of generators, labeled $X^i$, and 
$P^i$, to maintain an analogy with the canonical case. The generators obey 
the following commutation relations,
\beq
[P^j,X^i] = \delta _{i,j}\theta _i(X^i) \label{poisson1},
\eeq
and
\beq
[X^i,X^j] = [P^i,P^j] = 0.
\eeq
$\theta _i(X^i)$ (no sum over the repeated index $i$) in (\ref{poisson1}) 
corresponds to an arbitrary function 
of 
$X^i$. 
After a suitable choice of ordering is provided, the expression for the  
deformed product of functions in the non-commutative space of interest to
us (\ref{poisson1}) becomes,
\beq
f(x,p) \st g(x,p) = f(x,p) \exp \lb \sum_{i}
\stackrel{\leftarrow}{\dd{}{p^i}}\stackrel{\longrightarrow}{\lb\theta
^i(x^i)\dd{}{x^i}\rb} \rb g(x,p) \label{starproduct}.
\eeq

This class of non-commutative spaces include several physically 
interesting examples. The space whose coordinates obey the canonical 
commutation relations (\ref{canonical}), corresponds to the special case 
$\theta (X) =$ constant. The two dimensional $\kappa $-Minkowski space 
\cite{kmin1, kmin2}, 
also falls in this category. In this non-commutative space, 
the commutation relation among the 
coordinates $T$, $X$ is,
\beq
[T,X] = -\frac{i}{\kappa }X \label{min}.
\eeq
This space is obviously of the type considered in (\ref{poisson1}), with 
$T$ assuming the role of $P$, and $\theta (X)$ becoming a linear function 
of $X$. The methods used  for calculating the star 
product can easily be extended to the four dimensional $\kappa $-Minkowski 
space as well, and the corresponding result is quoted in the next section.

The h-deformed plane \cite{hplane1, hplane2, hplane3, hplane4} is another 
example of a non-commutative space that 
falls in this category. This space is the algebra generated by two 
generators $P$, and $X$, satisfying,
\beq
[P,X] = hX^2.
\eeq
A detailed proof of the expression for 
the star product (\ref{starproduct}), and the analysis of gauge theories 
using this star product is provided in the following sections.

\section{The Star Product}
For the purpose of simplifying the notation and  computations, we shall 
work with the two dimensional analog of (\ref{poisson1}) in the rest of 
the paper; i.e. the case of two non-commuting coordinates $X$ and $P$. 
The results can be generalized to  spaces made of many copies of this 
two 
dimensional space in a straightforward manner.

To set up an isomorphism between functions of 
non-commuting objects $(X,P)$, 
and functions of ordinary commuting variables $(x,p)$, one requires a 
rule, 
$(\Omega )$, for  associating a unique function of non-commuting 
variables to a 
given function of the ordinary commuting ones. i.e.
\begin{equation}
\Omega : f(x,p) \stackrel{\Omega}{\rightarrow} F(X,P) = \Omega 
(f(x,p) 
\end{equation} 
This isomorphism corresponds to a quantization. In the present paper we 
seek to quantize a Poisson manifold with the following Poisson bracket.
\beq
\{p,x\} = \theta (x) \label{classicalspace}
\eeq
This Poisson bracket is coordinate dependent. But, as long as there exists 
some coordinate system in which the Poisson bracket can be brought to this 
form, the analysis goes through. 

We shall think of the functions $f(x,p)$ 
of the commuting variables as formal power series of the form,
\begin{equation}
f(x,p) = \sum_{m,n}a_{m,n}x^mp^n.
\end{equation}
The functions of the non-commuting variables $F(X,P)$ will also be thought 
of as elements of the space of formal power series generated by the two 
letters $X,P$, modulo the relations that characterize the particular 
non-commutative space of interest. 
In the non-commutative space of interest to us, the relations 
between the two coordinates (denoted by $P$ and $X$) are,
\begin{equation}
[P,X]=\theta(X) \label{alg}
\end{equation}
These commutation relations can be solved by representing $X$ by the 
multiplication operator $x$ and $P$ by $\theta(x)\frac{\partial}{\partial 
x}$.Assuming that $\theta (x)$ admits a power series expansion, we have,
\begin{equation}
P = \theta(x)\frac{\partial}{\partial x} =\sum_{n} \theta _n 
x^n\frac{\partial}{\partial x} \label{rep}
\end{equation}
In the general case (\ref{poisson1}) too we can solve the commutation 
relations as,
\beq
P^i = \theta ^i(x^i)\dd{}{x^i},
\eeq
where no sum is implied over the index $i$.

Specifying $\Omega $ amounts to choosing an ordering, and for the present 
problem, we shall chose the `standard ordering'; for which,
\begin{equation}
\Omega (x^mp^n) = X^mP^n = \Omega (p^nx^m)
\end{equation}

To complete the isomorphism between the functions of commuting variables 
and those of non-commuting ones, we need a $\Omega $ dependent deformation 
of the ordinary point-wise multiplication of functions of the commuting 
variables. The deformed product is defined as,
\begin{equation}
f(x,p) \st g(x,p) = \Omega ^{-1}(\Omega(f(x,p))\Omega(g(x,p))) \label{corr}
\end{equation}

To work out the star product corresponding to standard ordering, let us 
first consider standard ordered operator valued functions 
\beq
F = X^aP^b = \Omega (f(x,p)=x^ap^b),
\eeq 
and,
\beq 
G = X^rP^s = \Omega (g(x,p)=x^rp^s).
\eeq
These functions are elements of a `basis' for expanding an 
arbitrary standard ordered function as a formal power series. So the 
extension of the 
star product to arbitrary functions can be had once it is worked out for 
the basis elements.   

The product of two basis elements,
\begin{equation}
FG=X^aP^bX^rP^s = X^{a+r}P^{b+s} + X^a[P^b,X^r]P^s,
\end{equation}
can be written as a standard ordered function, if the commutation 
relation $[P^a,X^r]$ can be expressed in the standard form. This can 
indeed be done using the representation of $P$ as a differential operator 
described above. 

Using this representation, it can be shown that,

{\bf Proposition 1}
\begin{equation}
P^bX^r = \sum_{\stackrel{l=0}{n_1,...,n_l=0}}^{\stackrel{\infty }{b}} 
C_l(\theta,r,b,n_1,...n_{l-1})X^{(r+(n_1 
-1)+(n_2-1)...(n_l-1))}P^{b-l}, \label{lemma1}
\end{equation}
where,
\beq
C_l(\theta,r,b,n_1,...n_{l-1})=\chose{b}{l}\theta _{n_1}...\theta 
_{n_l}r(r+(n_1-1))...(r+(n_1-1)+...+(n_{l-1}-1)). \label{lemma2}
\eeq
In the expression for the coefficients $C_l(\theta,,r,b,n_1,...,n_{l-1})$ 
in the proposition above, no sum is implied over the repeated indices 
$n_1,...n_{l-1}$.
Hence,
\begin{eqnarray}
& &FG = X^aP^bX^rP^s \nonumber\\
& &= \sum_{l=0}^b 
C_l(\theta,r,b,n_1,...n_{l-1})X^{r+(n_1-1)...(n_l-1)+a}P^{b+s-l}\nonumber\\
& &
\end{eqnarray}
From the definition of the star product (\ref{corr}), it now follows that
\begin{equation} 
f\star g = \Omega ^{-1}(\Omega (f)\Omega (g))= \sum _{l=0}^b 
\frac{1}{l!}\left( \frac{\partial ^l}{\partial 
p^l}x^ap^b\right)\left( (\theta (x)\frac{\partial}{\partial 
x})^lx^rp^s\right).
\end{equation}

Extending this analysis to arbitrary differentiable functions $f,g$, we 
have.
\begin{equation}
f \star g = f e^{\stackrel{\leftarrow}{\frac{\partial}{\partial 
p}}\stackrel{\longrightarrow}{(\theta 
(x)\frac{\partial}{\partial x})}} g
\end{equation}
A straightforward generalization to algebras containing several 
pairs of the generators $P^i$ and $X^i$ and the relations 
(\ref{poisson1}) gives us (\ref{starproduct}). 

{\bf The case of the $\kappa $-Minkowski space:} As an aside, it is worth 
mentioning that a similar ordering prescription leads to a simple formula 
for the star product in the $\kappa $-Minkowski space. The four 
dimensional 
$\kappa $-Minkowski space is generated 
by the generators, $T$ and $X^i$, satisfying the following relations.
\beq
[T,X^i] = -\ovk X^i. \label{kmin2}
\eeq
The $X^i$'s commute among themselves. We pick `standard ordering' 
between  $X^i$ and $T$ as the 
preferred ordering prescription in this non-commutative space. 
Ordering among the $X$ variables does not matter as they commute with 
each other. More specifically,
\beq
\Omega \lb \prod _{i=1}^3 (x^i)^{r_i} t^j \rb = \Omega \lb t^j \prod 
_{i=1}^3 (x^i)^{r_i} \rb = \prod _{i=1}^3 (X^i)^{r_i}T^j.
\eeq
Some recent advances in the non-commutative geometry of $\kappa 
$-Minkowski spaces from a similar point of view can be found in 
\cite{kmin3, lizzi}.

A standard ordered function on 
the $\kappa $-Minkowski space, $F(T,X^1,X^2,X^3)$ can be thought of as a 
formal power series,
\beqs
F(T,X^1,X^2,X^3) &=& \sum _{r_1,r_2,r_3,i = 0}^{\infty} 
f_{r_1,r_2,r_3,r}(X^1)^{r_1}(X^2)^{r_2}(X^3)^{r_3}T^i \nonumber\\
& &= \Omega 
(f(t,x^i,x^2,x^3))
\eeqs
Once this choice of ordering is made, we can, 
as was done  in (\ref{lemma1}), bring
products of standard ordered functions to standard
forms by
commuting $T$ through the $X^i$'s using (\ref{kmin2}) repeatedly. It is 
tedious but  
straightforward to see that the resulting expression for star product in 
the $\kappa 
$-Minkowski space is,
\beq
(f \star g)(t,x^1,x^2,x^3,) = f(t,x^1,x^2,x^3) \prod _{j= 1}^3 
e^{\stackrel{\leftarrow}{\dd{}{t}}\lb \stackrel{\longrightarrow}{-\ovk x 
^j\dd{}{x^j}}\rb} g(t,x^1,x^2,x^3).
\eeq
In the argument of the exponential in the above equation, no sum is 
implied over the repeated index $j$. 

{\bf Proof of Proposition 1:}\\
We present a proof based on induction on the index $b$ in (\ref{lemma1}). 
For $b=1$, it is easy to see that,
\beq
PX^r = \sum_{n_1}\theta _{n_1}x^{n_1}\pd _xx^r = 
\sum_{n_1}\theta _{n_1}rX^{r+(n_1-1)} 
+ X^rP.
\eeq
Hence the formula holds for $b=1$.\\
Using the representation of $P$(\ref{rep}), it is again straightforward to 
verify that, for $b=2$,
\beqs
P^2X^r &=& \sum_{n_1,n_2} \theta 
_{n_1}\theta_{n_2}r(r+(n_1-1))X^{r+(n_1-1)+(n_2-1)} + 
\nonumber \\
& &2\sum_{n_1}\theta 
_{n_1}rX^{r+(n_1-1)}P + X^rP^2.
\eeqs
Hence the formula holds for $b=2$.
Assuming, that (\ref{lemma1}) hold for a particular value of $b$, one can 
use (\ref{rep}) once again to get,
\beqs
&&P^{b+1}X^r = \nonumber \\
&&\sum_{l}C_l(\theta,r,b,n_1,\dots, n_{l-1})\theta 
_{n_{l+1}}(r+(n_1-1)+\dots +(n_l-1))\times \\
&&\quad X^{r+(n_1-1)+\dots 
+(n_{l+1}-1)}P^{b-l}
+\nonumber \\
&&\sum_{l}C_l(\theta,r,b,n_1,\dots, 
n_{l-1})X^{r+(n_1-1)+\dots (n_{l-1}-1)}P^{b-l+1}. 
\label{induction}
\eeqs
We now note that the coefficients satisfy the following recursion relation,
\beqs
C_t(\theta,r,r,n_1,...,n_{t-1})\theta_{n_{t+1}}(r+(n_1-1)+...+(n_t-1)) + 
\\
C_{t+1}(\theta,r,b,n_1,...n_t) = C_{t+1}(\theta,r,b+1,n_1,...n_t).
\eeqs
Hence, by combining the coefficients for each value of $l: (0<l<b)$ from 
the second 
series in (\ref{induction}) with that of the $l-1$th term from the first 
one, we obtain,
\beq
P^{b+1}X^r = 
\sum_{l=0}^{b+1}C_l(\theta,r,b+1,n_1,...,n_{l-1})X^{r+(n_1-1)+...(n_l-1)}P^{b+1-l}
\eeq

\section{Construction of Gauge theories, and the Seiberg-Witten map}
To construct a gauge theory on the non-commutative space discussed above, 
we shall use the approach of Madore et al \cite{Madoreetal}. We shall now 
summarize some of 
their results in the context of a general non-commutative space 
coordinatized by the non-commuting variables $X^i$, obeying the relation,
\beq
[X^i,X^j]=\theta ^{ij}(X).
\eeq
In this approach, the fields $\phi$, are taken to be elements of the space 
of 
formal power series generated by $X^i$, modulo the defining relations 
given above. The effect of a gauge transformation on the fields is taken 
to be of the 
form,
\beq
\delta _{\alpha} \phi(X) = i\alpha (X)\phi(X).
\eeq
The coordinates $(X^i)$ themselves are taken to be invariant under the 
gauge transformations. Since left multiplication by a coordinate is not a 
gauge covariant operation, generalized coordinates,
\beq
Q^i = X^i + A^i  
\eeq
are introduced, and it is required that left multiplication by the 
generalized coordinates be a gauge covariant operation. i.e
\beq
\delta _{\alpha (X)}Q^i\phi (X) = i\alpha (X)Q^i\phi(X).
\eeq
This restriction implies that,
\beq
\delta _{\Al} A^i(X) = i[\Al,A^i(X) + X^i]. \label{gtransform}
\eeq
One can also construct a tensor $T^{ij}$:
\beq
T^{ij} = [Q^i,Q^j] -i\theta ^{ij}(X) = [A^i,X^j] + [X^i,A^j] +[A^i,A^j]
\eeq
This tensor is gauge-covariant, and satisfies,
\beq
\delta _{\Al}T^{ij} = i[\Al,T^{ij}].
\eeq
Clearly, $A^i(X)$ and $T^{ij}(X)$ are the generalizations of the 
Yang-Mills connection and curvature to  
non-commutative 
geometry. 

One can now  use the $\Omega $ correspondence(\ref{corr}) to translate 
the statements 
made above into statements about functions of commuting variables $x^i$ 
as follows.
\begin{eqnarray}
\de \phi (x) = [\al \st \phi (x)]\nonumber \\
\de A^i(x) = [\al \st (A^i(x) +x^i)] \nonumber\\
T^{ij} = [A^i(x) \st A^j(x)] + [x^i \st A^j(x)] + [A^i(x)\st x^j]\nonumber 
\\
\de T^{ij}(x) = i[\al \st T^{ij}(x)
\end{eqnarray}

In fact more is possible. Just as the algebra of operator valued functions 
can be thought of as the algebra of functions of commuting variables with 
a deformed product, there exists a map between gauge theories on 
commutative spaces and gauge theories on non-commutative ones. 
This is the Seiberg-Witten map \cite{Seiberg-Witten, 
Madoreetal, jurco1,jurco2, 
jurco3}. 

Denoting the connection on commutative spaces by 
$a_i(x)$, and the infinitesimal gauge parameter by $\epsilon (x)$, we 
recall that the gauge transformation is given by
\beq
\delta _{\epsilon (x)}a_i(x) = \pd _i\epsilon (x) + [\epsilon (x),a_i(x)].
\eeq
The connection $A_i(x)$ and the gauge parameter $\alpha (x)$ on the 
non-commutative space are  now to be viewed as non-linear functions of 
$a_i(x)$, $\epsilon (x)$ and $\theta ^{ij}(x)$ such that the law for gauge 
transformation in the non-commutative world (\ref{gtransform}) follows 
from the ordinary 
gauge transformation for $a_i(x)$. 

We now present an explicit construction of the Seiberg-Witten map (up to 
$O(\theta ^2)$ for the non-commutative space defined by (\ref{alg}). We 
shall 
concentrate on the abelian gauge theories in this paper. For this 
specific case, we have,
\beqs
[f \st g] &=& \ab \lb \dd{f}{p}\dd{g}{x} -\dd{g}{p}\dd{f}{x} \rb + 
\nonumber\\
& &\half \ab ^2 \lb \dd{^2f}{p^2}\dd{^2g}{x^2} - 
\dd{^2g}{p^2}\dd{^2f}{x^2}\rb + \nonumber \\
& &\half \ab \dd{\ab }{x}\lb \dd{^2f }{p^2}\dd{g}{x} - 
\dd{^2g}{p^2}\dd{f}{x}\rb + O(\ab ^3).
\eeqs

This allows us to write in an explicit form the variation of the 
connection under the `star' gauge transformations(\ref{gtransform}). These are given by,
\beqs
\de A_x &=& i\ab\lb \dd{\all}{p}\lb1+\dd{A_x}{x}\rb - 
\dd{A_x}{p}\dd{\all}{x}\rb + \nonumber \\
& &i\half\ab ^2\lb\dd{^2\all}{p^2}\dd{^2A_x}{x^2} 
-\dd{^2A_x}{p^2}\dd{^2\all}{x^2}\rb+\nonumber \\
& &i\half \ab \dd{\ab}{x}\lb \dd{^2}{p^2}\lb1+\dd{A_x}{x}\rb - 
\dd{^2A_x}{p^2}\dd{\all}{x}\rb + O(\ab ^3), \label{varA1}
\eeqs
and,
\beqs
\de A_p &=& i\ab\lb\dd{\all}{p}\dd{A_p}{x} - 
\lb1+\dd{A_p}{p}\rb\dd{\all}{x}\rb + \nonumber \\
& & i\half \ab^2\lb\dd{^2\all}{p^2}\dd{^2A_p}{x^2} - 
\dd{^2A_p}{p^2}\dd{^2\all}{x^2}\rb + \nonumber \\
& & i\half \dd{\ab}{x}\lb \dd{^2\all}{p^2}\dd{A_p}{x} - 
\dd{^2A_p}{p^2}\dd{\all}{x}\rb + O(\ab ^3). \label{varA2}
\eeqs
We now make the following ansatz,
\beqs
\all = \epsilon + \gamma (\ab,\epsilon,a) + O(\ab ^2) \nonumber \\
A_x = \ab a_p +g_x(\ab,a) + O(\ab^3)\nonumber \\
A_p = -\ab a_x +g_p(\ab,a) +O(\ab^3)
\eeqs
This ansatz guarantees that to the first order on $\ab$, the variation of 
$A$ is obtained by the variation of $a$. It is implied that $\gamma $ is 
of $O(\ab )$ and that $g_x,g_p$ are both of $O(\ab ^2)$. To go beyond the 
leading order, we need to put the ansatz in equations (\ref{varA1}) and 
(\ref{varA2}). This 
gives us expressions for the variations of $g_x$ and $g_p$; which are,
\beqs
\de g_x &=& i\ab ^2\lb\dd{\epsilon}{p}\dd{a_p}{x} - 
\dd{a_p}{p}\dd{\epsilon}{x}\rb + \nonumber \\
& &\ab\dd{\ab}{x}\lb\dd{\epsilon}{p}a_p +\half\dd{^2\epsilon}{p^2}\rb + 
\nonumber \\
& &i\ab\dd{\gamma}{p} + O(\ab^3),
\eeqs
and
\beqs
\de g_p &=& i\ab ^2 \lb \dd{\epsilon}{x} \dd{a_x}{p} - \dd{\epsilon}{p} 
\dd{a_x}{x} \rb - \nonumber \\
& & i\ab \dd{\ab}{x} \lb \dd{\epsilon }{p}a_x \rb -\nonumber \\
& & i \ab \dd{\gamma}{x}.
\eeqs
If we now allow  $\gamma $ to have the following form,
\beq
\gamma = \ab \dd{\epsilon }{x}a_p = \ab (\dee a_x)a_p, \label{var1}
\eeq
it then follows that,
\beqs
\de g_x = \dee \lb \half i\ab \dd{\ab}{x} \lb a_p^2 + \dd{a_p}{p}\rb + 
i\ab ^2\lb 
a_p \dd{a_p}{x}\rb \rb ,\nonumber \\
\de g_p = \dee \lb -i\ab ^2 \lb a_p \dd{a_x}{x}\rb -i\ab \dd{\ab}{x}\lb 
a_xa_p\rb \rb. \label{var2}
\eeqs
In deriving these expressions we have used the relation,
\beq
\dee \lb\dd{a_x}{p} - \dd{a_p}{x}\rb = \dd{^2 \epsilon }{x \pd p} - \dd{^2 
\epsilon }{x \pd p} = 0.
\eeq
Equations (\ref{var1}) and (\ref{var2}) provide us with the following explicit 
expressions for the 
Seiberg-Witten map at the first non-trivial order, 
\beq
\alpha = \epsilon +  \ab \dd{\epsilon }{x}a_p +O(\theta (x)^2).
\eeq
\beq
A_x = \theta (x)a_p + \lb \half i\ab \dd{\ab}{x} \lb a_p^2 + 
\dd{a_p}{p}\rb +
i\ab ^2\lb
a_p \dd{a_p}{x}\rb \rb +O(\theta (x)^3).
\eeq
\beq
A_p = -\theta (x)a_x + \lb -i\ab ^2 \lb a_p \dd{a_x}{x}\rb -i\ab 
\dd{\ab}{x}\lb
a_xa_p\rb \rb + O(\theta (x)^3).
\eeq 

{\bf Acknowledgment:} We would like to thank Prof S.G.Rajeev for his 
encouragement and support. We are also grateful to A.Constandache 
and G.S.Krishnaswami for many useful discussions.

\bibliography{abhishekbib}

\begin{thebibliography}{10}

\bibitem{Madoreetal}
J.Madore, S.Schraml, P.Schupp, and J.Wess,
\newblock Eur. Phys. J. {\bf C16}, 161 (2000).

\bibitem{Seiberg-Witten}
N.Seiberg and E.Witten,
\newblock JHEP {\bf 9909}, 032 (1999).

\bibitem{ncreview1}
A.Konechny and A.Schwarz,
\newblock Phys. Rept {\bf 360}, 353 (2002).

\bibitem{ncreview2}
R.J.Szabo,
\newblock hep-th/0109162.

\bibitem{Kontsevich}
M.Kontsevich,
\newblock q-alg/9709040.

\bibitem{Moyal}
J.Moyal,
\newblock Proc. Camb. Phil. Soc {\bf 45}, 99 (1949).

\bibitem{Bayenetal}
F.Bayen, M.Flato, M.Fronsdal, C.Lichnerowicz, and D.Sternheimer,
\newblock Ann. Phys {\bf 111}, 61 (1978).

\bibitem{connesbook}
A.Connes,
\newblock {\em Noncommutative Geometry} (San Diego: Academic Press, 1994).

\bibitem{groenwold}
H.J.Groenwold,
\newblock Physica {\bf 12}, 405 (1946).

\bibitem{Catfel}
A.S.Cattaneo and G.Felder,
\newblock Commun. Math. Phys {\bf 212}, 591 (2000).

\bibitem{BakasKakas}
I.Bakas and A.C.Kakas,
\newblock Class. Quantum Grav {\bf 4}, L67 (1987).

\bibitem{WolfAgarwal}
G.Agarwal and E.Wolf,
\newblock Phys. Rev. D {\bf 2}, 2187 (1970).

\bibitem{Dunne}
G.V.Dunne,
\newblock J. Phys. A {\bf 21}, 2321 (1988).

\bibitem{zachos1}
C.Zachos,
\newblock J.Math.Phys {\bf 41}, 5129 (2000).

\bibitem{zachos2}
C.Zachos,
\newblock hep-th/0008010.

\bibitem{kmin1}
S.Majid and H.Ruegg,
\newblock Phys. Lett. B {\bf 334}, 348 (1994).

\bibitem{kmin2}
J.Likierski, H.Ruegg, and W.J.Zakrzewski,
\newblock Ann. Phys {\bf 143}, 90 (1995).

\bibitem{hplane1}
Y.Manin,
\newblock {\em Topics in Noncommutative Geometry} (Princeton University Press,
  Princeton, 1991).

\bibitem{hplane2}
S.Cho, J.Madore, and K.Park,
\newblock J. Phys. A: Math. Gen {\bf 31}, 2639 (1998).

\bibitem{hplane3}
E.Demidov, Y.Manin, E.Mukhin, and D.Zhdanovich,
\newblock Prog. Theor. Phys (Suppl) {\bf 102}, 203 (1990).

\bibitem{hplane4}
J.Madore and H.Steinacker,
\newblock J. Phys. A: Math. Gen {\bf 33}, 327 (2000).

\bibitem{kmin3}
G.Amelino-Camelia and M.Arzano,
\newblock Phys. Rev. D {\bf 65}, 084044 (2002).

\bibitem{lizzi}
A.Agostini, F.Lizzi, and A.Zampini,
\newblock Mod. Phys. Lett. A {\bf 17}, 2105 (2002).

\bibitem{jurco1}
B.Jurco, P.Schupp, and J.Wess,
\newblock Nucl. Phys. B {\bf 584}, 784 (2000).

\bibitem{jurco2}
B.Jurco and P.Schupp,
\newblock Eur. Phys. J. C {\bf 14}, 367 (2000).

\bibitem{jurco3}
B.Jurco, P.Schupp, and J.Wess,
\newblock Nucl. Phys. B {\bf 604}, 148 (2001).

\end{thebibliography}

\end{document}